\documentclass[aps,prl,superscriptaddress,twocolumn]{revtex4}
\usepackage{bm}
\usepackage{graphicx,amsmath,latexsym,amssymb}
\usepackage{color}
\usepackage{dcolumn}

\begin{document}

\title{Transverse Momentum Transfer in Atom-Light Scattering }

\author{B.A. van Tiggelen}

\affiliation{Universit\'{e} Grenoble 1/CNRS, LPMMC UMR 5493, B.P.
166, 38042 Grenoble, France}

\author{A. Nussle}

\affiliation{Universit\'{e} Grenoble 1/CNRS, LPMMC UMR 5493, B.P.
166, 38042 Grenoble, France}

\author{G.L.J.A. Rikken}

\affiliation{LNCMI, UPR 3228 CNRS/INSA/UJF Grenoble 1/UPS, Toulouse
\& Grenoble, France }

\date{\today}

\begin{abstract}
We predict a photon Hall effect in the optical cross-section of
atomic hydrogen, which is caused by the interference between an
electric quadrupole transition and an electric dipole transition
from the ground state to $3D_{3/2}$ and $3P_{3/2}$. This induces a
magneto-transverse acceleration comparable to a fraction of  $g$. In
atoms with a two level electric dipole transition, a much smaller
transverse force is generated only when the atom is moving.
\end{abstract}


\maketitle

Light scattering exchanges momentum between matter and radiation,
and thus induces a force on the matter.
Classical light scattering is well known to be affected by a
magnetic field. A specific feature, the photon Hall effect (PHE),
was first predicted in multiple light scattering \cite{prlbart}, and
observed shortly afterwards \cite{naturegeert} with typical changes
in the magneto-transverse photon flux of order $ 10^{-5}\, $ per
Tesla of applied magnetic field. A Mie theory for the PHE
\cite{JOSADavid} agreed quantitatively with the experiments. Given
the wave number $\mathbf{k}$ of the incident photon flux and the
magnetic field ${\mathbf{B}}$ , the PHE induces an exchange of
momentum between scatterer and radiation in the magneto-transverse
(''upward") direction along $ \mathbf{B}\times \mathbf{k}$. A light
flux of $10^4$ W/m$^2$ incident on a micron-sized particle with a
relative PHE of $10^{-5}$ per Tesla experiences a transverse force
of $10^{-19}$ N/T, roughly equivalent to the Lorentz force on a
charge $e$ moving with a velocity of $1$ m/s. The magneto-transverse
acceleration for a  $10 \, \mu$m TiO$_2$ particle would be as small
as $10^{-11}$  m/s$^2$ in a field of $10$ Tesla.

Atoms are strong light scatterers that can achieve  elastic optical
cross-sections as large as the maximum unitary limit $\lambda^2$
near optical transitions, and with promising applications in
mesoscopic physics \cite{cord}. When the typical Zeeman splitting
$\frac{1}{2} \omega_c$ ($\omega_c = e B /m_e = 17.5$ MHz/Gauss is
the cyclotron circular frequency) equals the atomic line width
(typically $\gamma \approx 100 $ MHz), the optical cross-section is
significantly altered by the magnetic field, typically true for a
few Gauss. Since atoms have small mass, the magneto-transverse
recoil would be much larger than for Mie particles. The
magneto-cross-section of an atomic resonance with width $\gamma$ and
Zeeman splitting $\omega_c$ can be estimated as $\frac{1}{2}
(\omega_c/\gamma)\lambda^2/\pi^2$. If we would assume  the Hall
cross-section to be  of this order the magneto-transverse
acceleration would be as large as $4$ km/s$^2$ per Gauss when tuned
the 1S-2P transition in Strontium exposed to a small flux of $100$
W/m$^2$. Unfortunately, no PHE can occur for pure electric-dipole
(ED) transitions, since the ED imposes a symmetry between forward
and backward scattering, as well as between upward and downward
directions in the magneto-cross-section \cite{JOSADavid}. The PHE
induced by the scattering from pairs of atoms in a cold $^{88}$Sr
gaz, is estimated to be a few percent \cite{Benoitprl}.

\begin{figure}[b]
\begin{center}$
\begin{array}{ccc}
\includegraphics[width=4.3cm]{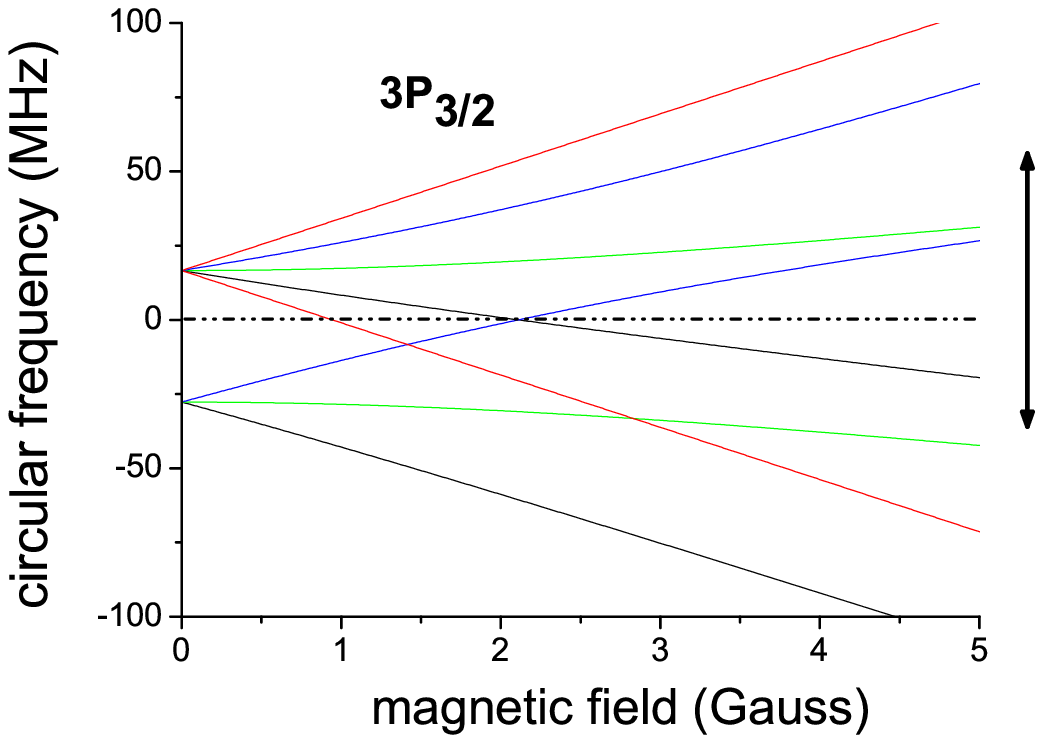} &
\includegraphics[width=4.3cm]{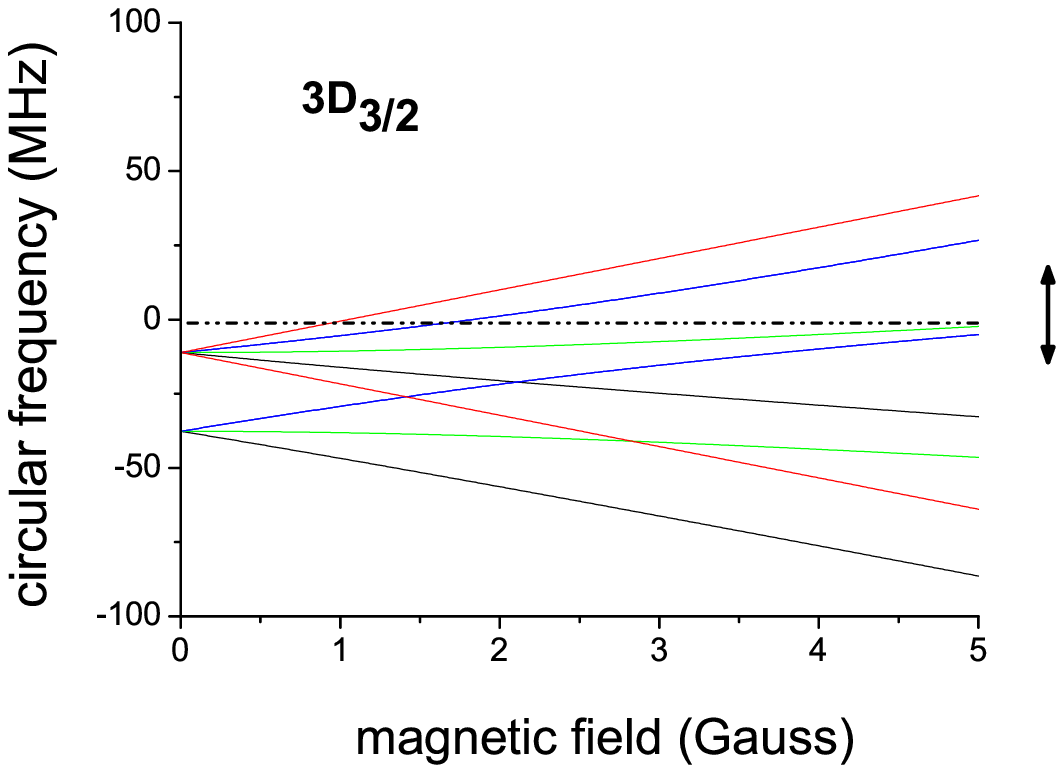} &
\end{array}$
\end{center}
\caption{Hyperfine structure of the $3P_{3/2}$ (left) and $3D_{3/2}$
(right)  level of atomic hydrogen, as a function of magnetic field.
Equal colors indicate equal values for the hyperfine magnetic
quantum number $m$. The height of the vertical bar on the right
indicates the line width $\gamma$. The zero in frequency is chosen
at the fine structure level of $3P_{3/2}$. The one of $3D_{3/2}$ is
$21.07$ radMHz lower.}\label{HF}
\end{figure}

Can the PHE of a single atom exist at all, and how large will the
magneto-transverse momentum transfer to the atom be?  Two striking
differences exist between classical Mie scattering and light-atom
scattering. First, given a monochromatic incident laser beam, the
atom is usually subject to inelastic transitions to levels that are
no longer excited by the same beam, thus preventing a stationary
scattering process. Secondly, given the small mass of atoms, one
must anticipate significant velocity recoils that change the
resonant frequency via the Doppler effect, and finally reduce the
light scattering.

The optical cross-section of an atom is expressed by the
Kramers-Heisenberg formula \cite{loudon},

\begin{eqnarray}\label{diffc}
  \frac{d\sigma}{d\Omega}\left( \omega\mathbf{k\varepsilon}\rightarrow \omega_s\mathbf{k}_{s}\mathbf{\varepsilon}_{s} \right)
  = \alpha^{2} \frac{\omega_{s}^{3}}{\omega^{3}}   \left| f_{ED} (\omega,\varepsilon,\varepsilon_s
  ) \right.\nonumber \\
\left.   + f_{EQ}
(\omega,\varepsilon,\varepsilon_s,\mathbf{k},\mathbf{k}_s ) + \cdots
\right|^2 \nonumber
\end{eqnarray}
Here, $\alpha$ is the fine structure constant, $\omega$ and
$\omega_s < \omega$ are incident and scattered frequency,
$\varepsilon$ and $\varepsilon_s$ are the polarization vectors of
incident and scattered radiation;  $f(\omega)$ is the complex
scattering amplitude associated with transitions in the atom, that
can be either elastic or inelastic, and driven by either electric
dipole (ED) or quadrupole (EQ). The above expression does not take
into account stimulated emission (SE). For this to be true we
require that $ {W(\omega_s,\mathbf{k}_s,\varepsilon_s)}<
{W_0(\omega_s)}$, with $W$ the radiation density  per steradian, per
bandwidth, per polarization, and $W_0 = \hbar \omega_s^3/(2\pi
c_0)^3 $ its value for the quantum vacuum.

We will first focus on the simplest atom, atomic hydrogen, whose
physics in a magnetic field has been studied in great detail
\cite{dd,jook}. This atom has the \emph{unique} property that the
fine-structure levels $3P_{3/2}$ and $3D_{3/2}$  strongly overlap,
despite their hyperfine structure (HF), The anomalous Zeeman effect
of the latter is shown in Fig.~1. For not too large magnetic fields
all levels are energetically close and can thus interfere
constructively. It is instructive to first simply ignore the spin of
both electron and proton, and to adopt a simple $1S$ ground state
and excited levels $3P$ and $3D$ separated by the (fine-structure
energy) of $18$ radHz ($=2\pi s^{-1}$). The electronic transitions
$1S \rightarrow 3P \rightarrow 1S$ and $1S \rightarrow 3D
\rightarrow 1S$ are now both elastic. The $1S-3D$ transition,
however, is ED forbidden and requires an EQ transition. The ED
transition between the ground state $1S$ and the $3P$ level reads
\begin{widetext}
\begin{eqnarray}\label{fed}
  f_{ED} (\omega) = \frac{\omega^2}{c_0} \sum_{m=0,\pm 1}
  \frac{ \{\langle 1S | \mathbf{r} | 3P_m\rangle \cdot
  \varepsilon_s\}
\{\langle 3P_m | \mathbf{r} | 1S  \rangle \cdot \varepsilon \}}
  {\omega-\omega_m^P(B) +i\gamma_P}\equiv \frac{\omega^2 r^2_{3P}}{c_0}
  \left[P_0(\varepsilon_s\cdot\varepsilon) +[P_1-P_0](\varepsilon_s\cdot
  \hat{\mathbf{z}})(\varepsilon\cdot \hat{\mathbf{z}}) + P_2(\omega) i \varepsilon_s\cdot (\varepsilon \times
  \hat{\mathbf{z}})\right] \nonumber \\
\end{eqnarray}
\end{widetext} Here $\left< i \right|
\mathbf{r}\left| j \right>$ is the ED matrix element between states
$i$ and $j$. It depends on orbital momentum but has constant radial
part $r_{3P}= 0.517 a_0$. The second expression is obtained by
inserting the orbital eigenfunctions, and by putting $P_i =
1/(\omega - \omega_m(i,B) +i\gamma_{P})$. In our simplified picture,
the Zeeman effect behaves normally ($m\omega_c/2 $). We choose
$\hat{\mathbf{k}} = \hat{\mathbf{x}}$, $\hat{\mathbf{B}} =
\hat{\mathbf{z}}$ and let $
 \hat{\mathbf{B}}\times \hat{\mathbf{ k }}= \hat{\mathbf{y}} $ be the Hall direction.
For the elastic EQ transition via the $3D$ level we find,
\begin{widetext}
\begin{eqnarray}\label{eq}
  f_{EQ} (\omega) = \frac{\omega^2}{c_0} \sum_{m=0,\pm 1, \pm 2}
  \frac{\{ \mathbf{k}_s
  \cdot \langle 1S |\frac{1}{2}\mathbf{r}\mathbf{r} | 3D_m \rangle \cdot
  \varepsilon_s\,\} \, \{
  \mathbf{k}\cdot \langle 3D_m |\frac{1}{2} \mathbf{r} \mathbf{r} | 1S \rangle \cdot \varepsilon
 \} }
  {\omega-\omega_m^D(B) +i\gamma_D}
\nonumber
\end{eqnarray}
\end{widetext} with $\gamma_D = 32 $ radMHz the natural line width of
the 3D level. This expression can again be developed by inserting
the orbital eigenfunctions associated with the $3D$ level,  at fixed
radial matrix element  $q_{3D}=0.867 a_0^2$.

The differential cross-section for incident unpolarized, broadband
light is obtained from the interference between the two transitions,
averaged over incident polarization, and summed over outgoing
polarization. The Hall terms are defined by the difference in flux
up and down along the vector $\hat{\mathbf{y}} $, and are all
characterized by a factor
 $i(\mathbf{\hat{k}}_s\cdot \hat{\mathbf{y}})$ that emerges in the cross-product of
 the two scattering amplitudes.
We shall write this as

\begin{eqnarray}\label{diffphe}
&&\frac{d\sigma}{d\Omega} =\alpha^2 \frac{1}{\Delta}\int_{\Delta}
d\omega \sum_{\varepsilon,\varepsilon_s} \mathrm{Re} \,
f^*_{ED}f_{EQ} = \frac{d\sigma^0}{d\Omega}\nonumber \\  &+&
\alpha^2\frac{\omega^6}{\Delta c_0^4} r_{3P}^2q_{3D}^2\mathrm{ Re}\,
i (\mathbf{\hat{k}}_s\cdot \hat{\mathbf{y}}) \sum_{m,m'}
\bar{F}_n(\omega,B)
A_{mm'}(\mathbf{k},\mathbf{\hat{k}}_s,\mathbf{\hat{B}}) \nonumber \\
\end{eqnarray}
The Hall cross-section is a sum over $3\times 5= 15$ cross-products
among the magnetic sublevels. The factor $\bar{F}_{mm'}$ is the
frequency-averaged cross-product of the complex line profiles. If
the bandwidth $\Delta$ largely exceeds the line widths then
\begin{eqnarray}\label{ave}
\bar{F}_{mm'}(\omega)&=&\int_{-\infty}^{\infty} {d \omega}
\frac{1}{\omega-\omega_P(m)-i\gamma_P}\frac{1}{\omega-\omega_D(m')+i\gamma_D}
\nonumber \\
&=&  \frac{{2\pi } i}{\omega_P(m)-\omega_D(m') +
i(\gamma_D+\gamma_P)}
\end{eqnarray}
  It can be seen that only 6
functions $A_{mm'}$ actually generate a PHE, with $ A_{0,m'=\pm 1
}=m' [1-2(\mathbf{\hat{k}}_s\cdot \mathbf{\hat{z}})^2]/60$ and
$A_{m= \pm 1,m'=\pm 2 }=[m+\frac{1}{2}m' +(-2m+\frac{1}{2}m')
   (\mathbf{\hat{k}}_s\cdot \mathbf{\hat{x}})^2 -\frac{1}{2}m'\frac{1}{2}(\mathbf{\hat{k}}_s\cdot
\mathbf{\hat{y}})^2]/60 $. Note that this simplified picture
highlights the PHE as a  ''which-way" event inside the hydrogen
atom. It is straightforward to calculate from Eq.~(\ref{diffphe})
the total magneto-transverse recoil force (black line in Fig. 2).

\bigskip

The present picture poses three problems. First we know that excited
$3P$ atoms will have a significant probability to decay
inelastically to the meta-stable state $2S$ so that the PHE process
would rapidly come to an end. Secondly, absorbed photons will
transfer momentum to the atom that will rapidly become Doppler
detuned from the incident laser. Finally, the inclusion of hyperfine
structure (HF) considerably complicates the above picture.

\begin{figure}[t]
\includegraphics[width=7cm]{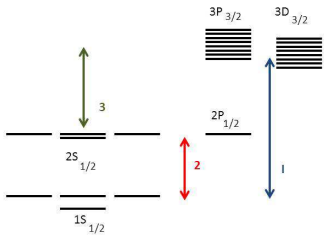}
\caption{Three different broadband laser beams are necessary to
generate a stationary process with magneto-transverse recoil. Laser
1 induces the transition to $3P_{3/2}$  and $3D_{3/2}$ levels that
generate the PHE. Laser 2 reassures that the inelastic decay to
$2S_{1/2}$ is pumped back to $3P_{3/2}$. Finally, beam 3,
propagating opposite to the beam 1 , compensates the longitudinal
photon recoil produced by laser 1.}\label{transitions}
\end{figure}

The longitudinal photon recoil to the atom of the first laser can be
compensated by a \emph{second} laser beam (intensity $I_2$) opposite
to the first, and exciting the atom to the $2P$ transition. If $I_2
\approx 3 I_1$ the average recoil rate is equal to zero (see Figure
2). The occurrence of inelastic decay to $2S$ must be compensated by
a \emph{third} laser beam (intensity $I_3$) that pumps  $2S$ atoms
back to $3P$. A straightforward analysis shows that detailed balance
results in $N_{2S}/N_{1S} = (\omega_{23}/\omega_{13} )^3 I_3/I_1$.
If the two laser intensities are roughly equal we infer that $N_{2S}
\ll N_{1S}$, so that the PHE with $1S$ as initial state is
maintained. Note that the $2S-(3P,3D)$ transitions also induce a PHE
which we will not discuss in view of its much smaller transverse
recoil.

\begin{figure}
\includegraphics[width=6cm]{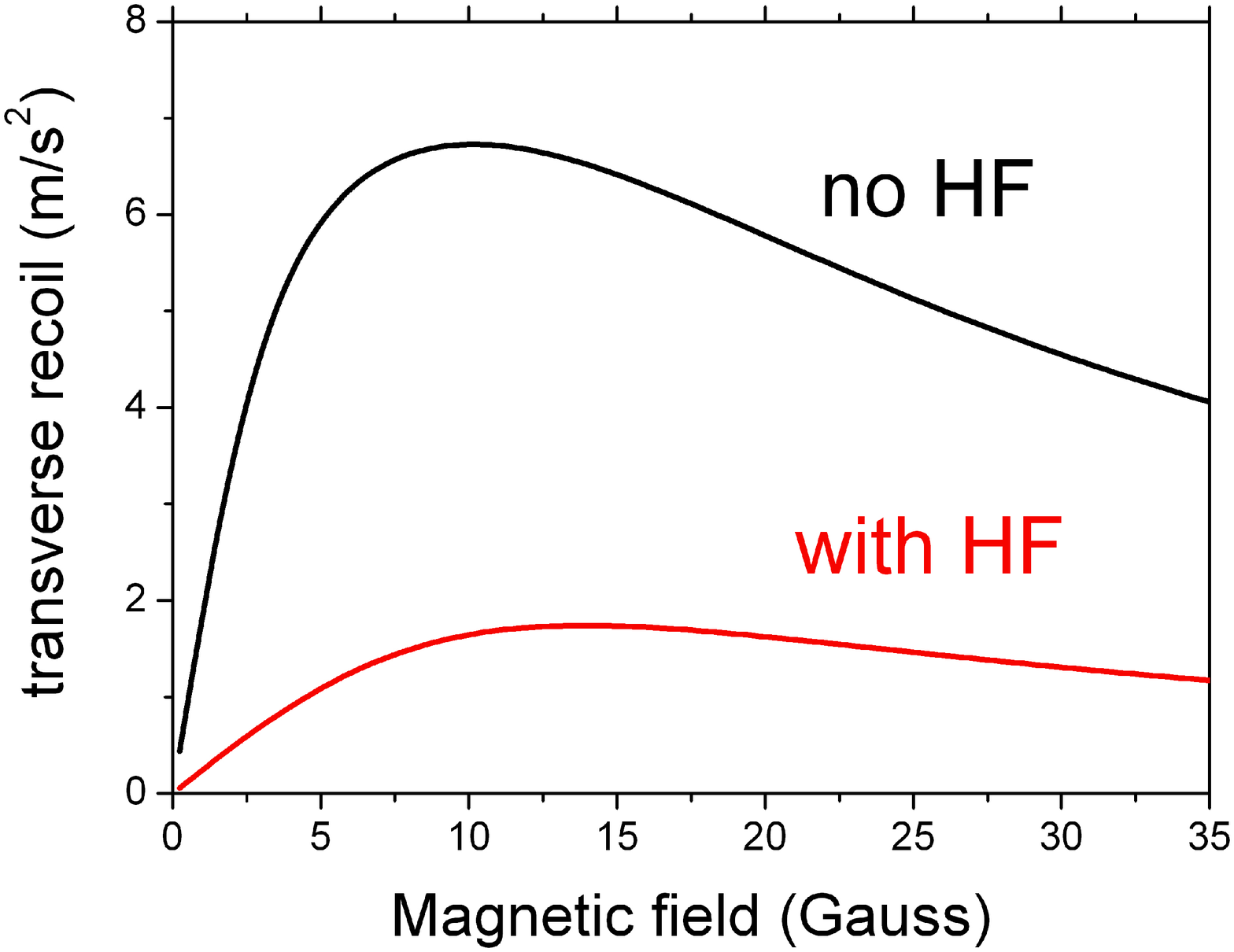}
\caption{Magneto transverse recoil acceleration of the hydrogen atom
in the presence of a broadband laser beam with flux $I= 10$ kW/m$^2$
and bandwidth $\Delta= 10 $ radGHz (thus resolving the HF ground
state). The black curve follows from the model with neglect of
(hyper)fine structure, the red curve takes into account the full
hyperfine structure of the $1S_{1/2}$ ground state and the
$3D_{3/2}$ and $3P_{3/2}$ levels, which results in a smaller PHE
recoil though with same sign. }\label{pherecoil}
\end{figure}

The inclusion of HF structure is  a straightforward process that we
shall not discuss in detail. The HF eigenfunctions
$\left|3P(D)_{j=\frac{3}{2}}, f= 1,2, m = -f,..,f \right>$ can be
constructed from the product states of orbital momentum, electron
and proton spin with appropriate Clebsch-Gordan coefficients. In the
presence of a magnetic field the magnetic sublevels $m=0, \pm 1$ of
the HF levels $f=1, 2$ mix, thus giving the $8$ sublevels whose
Zeeman effect is shown in Fig.~1. The $1S_{1/2}$ ground state splits
into one singlet and a triplet at $8.9$ radGHz higher in energy. The
PHE can be determined by collecting all cross products among the
transitions from the 4 $1S_{1/2}$ to the 8 $3D_{3/2}$ and $3P_{3/2}$
levels. The result of this cumbersome task is shown as the red line
in Figure 3. The HF splitting decreases the recoil because the
overlap between the levels decreases, roughly by a factor 4. In this
calculation it has been assumed that all HF levels are equally
populated, as a result of the presence of the two additional laser
beams and the broad band incident beam. In Figure 3 we show the
individual contributions of the 4 HF $1S_{1/2}$ states to the total
recoil. The spin-polarized states $|f=1, m=\pm 1\rangle$ have each a
nonzero, though opposite PHE at zero field that vanishes for equal
level population (a ``spin Hall effect", unobservable in the present
configuration with inelastic transitions that mix all up). In
addition, the PHE recoil from the unpolarized singlet state $|f=0,
m=0\rangle$ and the unpolarized triplet state $|f=1, m=0\rangle$ are
equal.

\begin{figure}
\includegraphics[width=6cm]{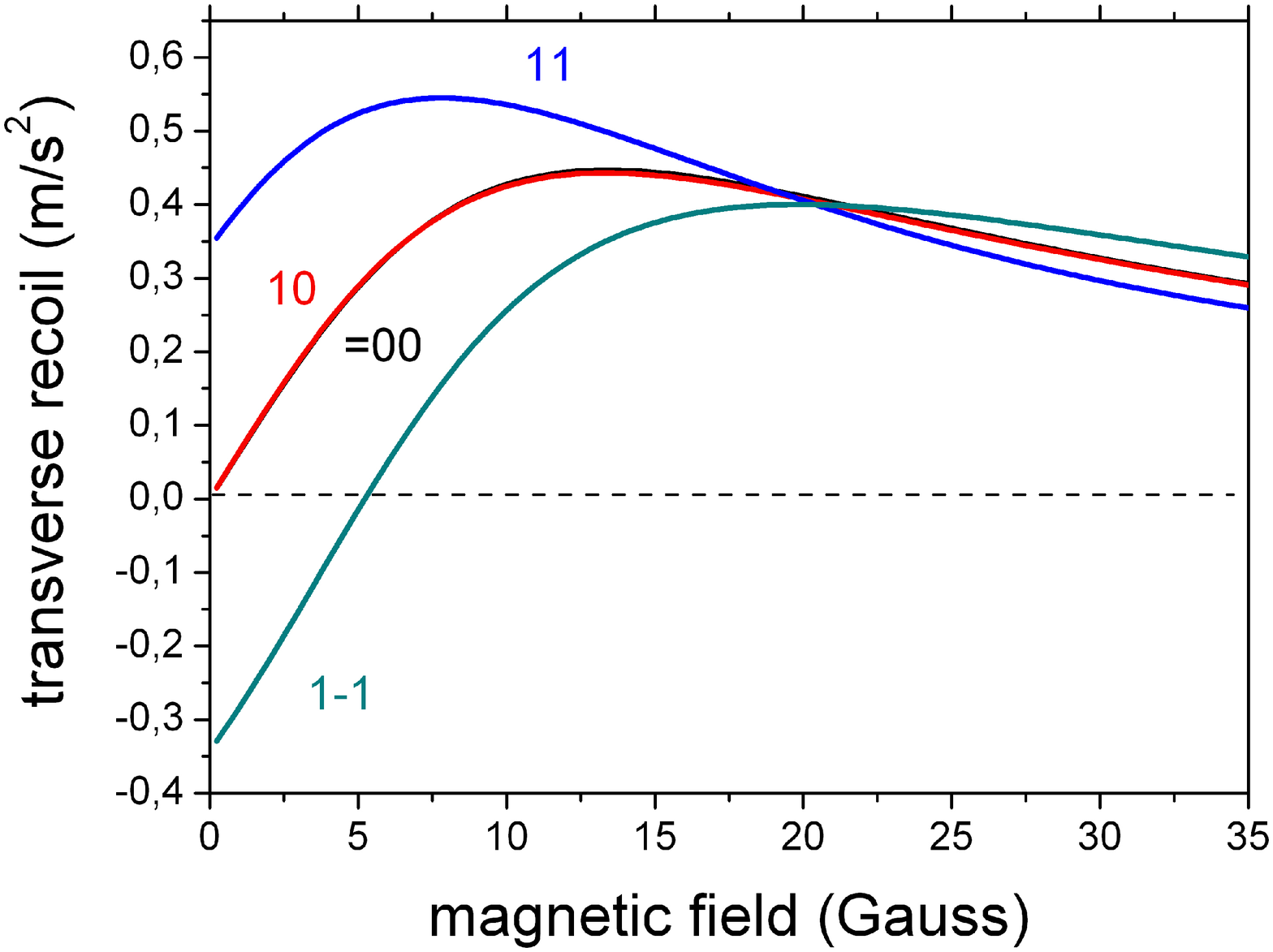}
\caption{Magneto transverse recoil of the hydrogen atom, as in
previous figure, here specified for transitions from the 4 HF ground
states $|f=0,1, m=-f,+f\rangle$. The red line in Fig 3 is the sum of
the 4. The curves for the two levels $f=0,1$, $m=0$  overlap.
}\label{pherecoilHF}
\end{figure}

\bigskip

We finally show that a PHE can be induced by an ED transition once
the atom is moving. Let $K'$ be the frame that moves with the atom,
and $K$ the one in which the atom moves with velocity $\mathbf{v}$.
The ED magneto-transverse scattering cross-section for a two-level
atom in frame $K'$ follows from the interference between first and
third term in Eq.~(1),
\begin{eqnarray}\label{diffED}
 \frac{d\sigma'}{d\Omega'}\left( \omega'\mathbf{k}'\rightarrow {\omega'}_s\mathbf{k}'_{s} \right)
 =  \frac{1}{9} {\omega'}_s^3\omega' \frac{\alpha^2r_{01}^4}{c_0^2}
 \mathrm{Im}(P_0P_2^*) \nonumber \\
\times  [\hat{\mathbf{k}}'_s\cdot \hat{\mathbf{k}}']
 [\hat{\mathbf{k}}'_s\cdot (\hat{\mathbf{B}}\times \hat{\mathbf{k}}')]
 \end{eqnarray}
This cross-section exhibits no net PHE since forward and backward
contribution cancel the flux along $\mathbf{k }\times \mathbf{B}$.
However, this cancelation is perturbed by the Doppler effect. The
cross-section $d \sigma$ relates an incident flux
$\rho(\omega,\mathbf{k})/c_0$ to an outgoing current
$\rho_s(\omega_s,\mathbf{k}_s) d \Omega r^2/c_0$ at a distance $r$
in the far field. Since $r$ is unaffected by a Lorentz
transformation to order $v/c_0$ , and since the radiation density
$\rho$ transforms as $\rho'= (\omega'^3/\omega^3)\rho $
\cite{milonni}, the cross-section in frame $K$ is
 \begin{eqnarray}\label{diffLor}
 \frac{d\sigma}{d\Omega}\left( \omega\mathbf{k}\rightarrow \omega_s\mathbf{k}_{s} \right)
   &\approx& \left(\frac{\omega' \omega_s}{\omega \omega'_s}\right)^3 \frac{d\sigma'}{d\Omega'}\left( \omega'\mathbf{k}'\rightarrow \omega'_s\mathbf{k}'_{s}
    \right) \nonumber
 \end{eqnarray}

We insert $\hat{\mathbf{k}}'_{(s)}
 \approx \hat{\mathbf{k}}_{(s)}(1+\hat{\mathbf{k}}_{(s)}\cdot \mathbf{v}/c_0) - \mathbf{v}/c_0$ and
 $\omega_{(s)}'\approx \omega_{(s)}(1-\mathbf{v}\cdot \hat{\mathbf{k}}_{(s)}/c_0)$ and assume for simplicity that the atom moves
 either parallel or opposite to the incident wave vector
 $\mathbf{k}$. Note that transformation factors involving $\omega'_s$ in the formula above
 cancel and do not contribute to the PHE.
The only contribution is the Lorentz transform of the
angle-dependent factor in Eq.~(\ref{diffED}) that
 generates $-[1-2(\hat{\mathbf{k}}_s\cdot \hat{\mathbf{k}})^2]
 (\mathbf{k}_s\cdot ({\mathbf{v}} \times \hat{\mathbf{B}})/c_0$.
The momentum transfer to the atom exhibits a magneto-transverse
force,
\begin{eqnarray}
 \mathbf{ F}&=& -\frac{1}{\hbar \omega}I(\mathbf{k}) \int d\Omega_s
 \, \hbar \mathbf{k}_s \frac{d\sigma}{d\Omega}\left( \omega\mathbf{k}\rightarrow
  \omega_s\mathbf{k}_{s} \right)  \nonumber \\ &=& I(\mathbf{k})\frac{4\pi }{45} \frac{\alpha^2 \omega^4 r_{01}^4 }{c_0^3}
  \mathrm{Im}(P_0P_2^*)(\omega\mp \frac{v\omega}{c_0}, B)\,  \frac{\mathbf{v}}{c_0}
    \times \hat{\mathbf{B}}\nonumber \\
\end{eqnarray}
with $I$ the incident flux in W/m$^2$. Apart from the Doppler shift
$\mp v/c_0$ of the incident frequency, this force is equal for
motion parallel ($-$) or opposite to the incident beam. This is
useful since the normal scattering of the light beam also induces a
longitudinal acceleration.  The presence of two opposite beams with
properly chosen intensities allows to induce a magneto-transverse
recoil and \emph{at the same time} select a specific constant
velocity. For typical parameters in a total flux of $100$ W/m$^2$
(small enough for stimulated emission to be small)  applied to
$^{88}$Sr we find the stable velocity $v=4.4$ m/s and a
magneto-transverse acceleration $a= 43$ $\mu$m /s$^2$ (Fig. 5). This
is much smaller than what we found for hydrogen, since $^{88}$Sr is
heavier and
 $v/c_0 \ll \alpha^2$, but is still measurable.

\begin{figure}
\includegraphics[width=7cm]{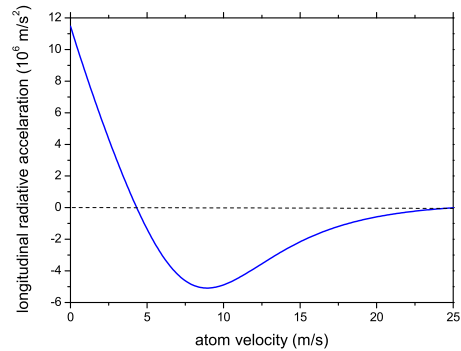}
\caption{Longitudinal acceleration of $^{88}$Sr by two opposite
beams with total flux ($I=100$ W/m$^2$), as a function of atom
velocity. The atomic transition is taken red-tuned by one line width
($\gamma= 101$ MHz) in a transverse magnetic field of $B=1$ Gauss.
The beam parallel to the speed is $50 \%$ more intense so that a
stable velocity $v=4.4$ m/s is selected (acceleration $a=0$).
}\label{move}
\end{figure}

In conclusion, we have quantified the magneto-transverse scattering
of light from unpolarized atomic hydrogen. It is caused by the
interference of en electric dipole transition and a electric
quadrupole transition. A transverse recoil of several m/s$^2$ is
predicted, i.e. a fraction of $g$. The generalization to other atoms
seems difficult since one needs overlapping transitions with
different (orbital) symmetry. these are often excluded by
(hyper)fine splitting. Maybe the application of high magnetic fields
may induce   level-crossing of remote transitions, this causing a
PHE. An electric dipole \emph{alone } is found to induce a
magneto-transverse scattering only when  the atom is moving, though
with much smaller accelerations of order $\mu$m/s$^2$ It could be
interesting to study the atomic spin-Hall effect in the
spin-polarized $S$-state of atomic hydrogen \cite{jook}, and to make
a link with previous predictions \cite{kwek}

 This work was supported by the ANR contract PHOTONIMPULS
ANR-09-BLAN-0088-01.

\end{document}